\documentclass[preprintnumbers,amsmath,amssymbm,preprint]{revtex4}
\usepackage{epsfig}
\usepackage{graphicx}

\begin{document}
\title{Survival probabilities in biased random walks: To restart or not to restart? that is the
question}
%\title{To restart or not to restart? This is the question}
%\title{Survival probabilities in biased random walk models: The Sisyphus versus the standard random walkers}
%\title{Survival probabilities in biased random walks with restarts}
\author{Shahar Hod}
\affiliation{The Ruppin Academic Center, Emeq Hefer 40250, Israel}
\affiliation{ }
\affiliation{The Hadassah Institute, Jerusalem 91010, Israel}
\date{\today}

\begin{abstract}
\ \ \ The time-dependent survival probability function $S(t;x_0,q)$
of biased Sisyphus random walkers, who at each time step have a
finite probability $q$ to step towards an absorbing trap at the
origin and a complementary probability $1-q$ to return to their
initial position $x_0$, is derived {\it analytically}. In
particular, we explicitly prove that the survival probability
function of the walkers decays exponentially at asymptotically late
times. Interestingly, our analysis reveals the fact that, for a
given value $q$ of the biased jumping probability, the survival
probability function $S(t;x_0,q)$ is characterized by a {\it
critical} (marginal) value $x^{\text{crit}}_0(q)$ of the initial gap
between the walkers and the trap, above which the late-time survival
probability of the biased Sisyphus random walkers is {\it larger}
than the corresponding survival probability of standard random
walkers.
\end{abstract}
\bigskip
\maketitle

\section{Introduction}

The phenomenology rich dynamics of biased random walkers
\cite{RW1,RW2,RW3,RW4,RW5,RW6} who are constantly drifting towards
an absorbing trap have been analyzed in diverse physical,
biological, and social models \cite{DV3,DV5,DV1,DV4,DV2}. In
particular, the physically intriguing concept of a biased random
walk plays a key role in the theories of absorbing-state phase
transitions \cite{DV1} and complex adaptive systems \cite{DV2}, in
the physics of granular segregation \cite{DV3} and polymer
adsorption \cite{DV4}, and in biological models of epidemic
spreading \cite{DV5}.

It is well established that biased random walkers who, at each
discrete time step, have a probability $q>1/2$ to step towards an
absorbing trap (and a smaller probability $1-q$ to step in the
opposite direction of the trap) are characterized by exponentially
decaying asymptotic survival probabilities of the form
\cite{DV2,Notesur}:
\begin{equation}\label{Eq1}
{{S(t+1)}\over{S(t)}}\to 2\sqrt{q(1-q)}\ \ \ \ \text{for}\ \ \ \
t\gg1\  .
\end{equation}

In this paper we shall analyze the time-dependent survival
probabilities of biased random walkers in the so-called Sisyphus
random walk model \cite{RS1,RS2,RS3,RS4,RS5,Sisy,Hodanp}. As opposed
to standard random walkers, a Sisyphus random walker is
characterized by a restart mechanism \cite{Sisy} by which, at each
discrete time step, she has a non-zero probability to return to her
initial location $x_0>0$.

The main goal of the present paper is to study {\it analytically}
the influence of the Sisyphus restart mechanism on the survival
probabilities of the biased random walkers. In particular, we here
raise the following physically interesting question: Comparing the
survival probability of the biased Sisyphus random walker with the
corresponding survival probability of a standard random walker (who
has no restart mechanism), which biased walker has the larger chance
to survive for asymptotically long times?

Interestingly, our analysis, to be presented below, reveals that the
answer to the above stated question is highly non-trivial. In
particular, the Sisyphus restart mechanism, which randomly offers
the walkers the opportunity to return to their initial position
$x_0$, affects the time-dependent survival probabilities of the
walkers in two opposite ways:
\newline
(1) On the one hand, the restart mechanism, which is applied at each
discrete time step with probability $1-q$, may increase the safe gap
between the biased Sisyphus random walker and the absorbing trap to
its maximally allowed value $x_0$ \cite{Notextr}. The corresponding
survival probability of such lucky walkers is therefore expected to
be increased by the restart mechanism which characterizes the
Sisyphus random walk model.
\newline
(2) On the other hand, the existence of an upper bound $x(t)\leq
x_0$ on the gaps between the walkers and the absorbing trap (a bound
which does not exist in the standard random walk model) prevents the
Sisyphus random walkers from reaching the asymptotically safe region
$x\to\infty$. This effect is therefore expected to decrease the
survival probabilities of biased Sisyphus random walkers as compared
to the corresponding survival probabilities of standard random
walkers.

Intriguingly, in this paper we shall reveal the fact that the
asymptotic late-time survival function $S(t;x_0,q)$ of the biased
Sisyphus random walkers displays a kind of a phase transition. In
particular, using analytical techniques, we shall explicitly prove
that, for a given value $q$ of the biased jumping probability, there
exists a critical (marginal) value $x^{\text{crit}}_0(q)$ of the
initial gap between the walkers and the trap, above which the
survival probability of the biased Sisyphus random walkers is {\it
larger} than the corresponding survival probability (\ref{Eq1}) of
the standard random walkers.

\section{Description of the system}

We study the dynamics of a biased random walker in the presence of
an absorbing boundary. At each time tick the walker, who is located
on the non-negative integers ($x\in\{0,1,2,...\}$), has a
probability $q>1/2$ to take one step towards the absorbing trap
which is located at the origin ($x_{\text{trap}}=0$) and a
complementary probability $1-q$ to jump back to her initial position
$x_0>0$. That is,
\begin{equation}\label{Eq2}
x(t+1)=
\begin{cases}
x(t)-1& \text{with probability}\ \ q\ ;
\\ x_0 & \text{with probability}\ \ 1-q\  .
\end{cases}
\end{equation}

The dynamics of the biased Sisyphus random walkers is dominated by
the mathematically compact rule (\ref{Eq2}) with the initial
conditions
\begin{equation}\label{Eq3}
N_{\text{tot}}(t)=1 \ \ \ \ \text{for}\ \ \ \ t=0,1,...,x_0-1\  .
\end{equation}
Here
\begin{equation}\label{Eq4}
N_{\text{tot}}(t)=\sum_{k=1}^{k=x_0}N_k(t)\
\end{equation}
denotes the total number of random walkers who, after taking $t$
steps, have not been absorbed by the trap, where $N_k(t)$ is the normalized
time-dependent number of random walkers who are located, at time
$t$, at the discrete position $x(t)=k>0$.

\section{Survival probabilities in the biased Sisyphus random walk model}

In the present section we shall analyze the survival probabilities
$S(t;x_0,q)$ of the biased random walkers. The time-dependent function
[see the normalization (\ref{Eq3})]
\begin{equation}\label{Eq5}
S(t;x_0,q)\equiv N_{\text{tot}}(t)\
\end{equation}
denotes the fraction of biased random walkers who have not been
absorbed by the trap during the time interval $t$. This survival
function depends monotonically on the number $t$ of steps taken by
the walkers and on the initial gap $x_0$ of the walkers from the
absorbing trap. Our main goal is to determine analytically the
temporal and spatial functional behaviors of this fundamental
function.

The time-dependent number $d(t)$ of biased random walkers who are
absorbed by the trap after taking exactly $t$ steps is given by
\begin{equation}\label{Eq6}
d(t)=N_{\text{tot}}(t-1)-N_{\text{tot}}(t)\  .
\end{equation}
Taking cognizance of the jumping rule (\ref{Eq2}) and the
normalization (\ref{Eq3}), one finds that this number of absorbed
agents is given by \cite{Notefct1}
\begin{equation}\label{Eq7}
d(t)=q^{x_0}\cdot N_{x_0}(t-x_0)\  .
\end{equation}
Substituting the relation \cite{Notefct2}
\begin{equation}\label{Eq8}
N_{x_0}(t-x_0)=(1-q)\cdot N_{\text{tot}}(t-x_0-1)\
\end{equation}
into Eq. (\ref{Eq7}) and using the identity (\ref{Eq5}), one obtains
the discrete recurrence relation
\begin{equation}\label{Eq9}
S(t;x_0,q)=S(t-1;x_0,q)-q^{x_0}(1-q)\cdot S(t-x_0-1;x_0,q)\
%N_{\text{tot}}(t)=N_{\text{tot}}(t-1)-q^{x_0}(1-q)\cdot N_{\text{tot}}(t-x_0-1)\
\end{equation}
for the fraction of biased random walkers who have not been absorbed
by the trap until the $t$-th time step. It is important to emphasize
that Eqs. (\ref{Eq8}) and (\ref{Eq9}) are valid for $t>x_0$. In
particular, the survival probability function is characterized by
the relation
\begin{equation}\label{Eq10}
S(t=x_0;x_0,q)=1-q^{x_0}\
\end{equation}
at $t=x_0$. The characteristic relation (\ref{Eq10}) follows from
the jumping rule (\ref{Eq2}), according to which a biased random
walker who is initially located at $x_0$ has to go $x_0$ steps in a
row towards the trap (with a characteristic probability $q$ for each
step to the left) in order to be absorbed at the origin.

Our main goal is to analyze the asymptotic large-$t$ behavior of the
survival probability function. To this end, it proves useful to use
the mathematical ansatz
\begin{equation}\label{Eq11}
S(t;x_0,q)=\alpha(x_0,q)\cdot[\beta(x_0,q)]^t\  ,
\end{equation}
which, as we shall explicitly show below, is valid in the asymptotic
large-$t$ regime (see the data presented in Table \ref{Table1}
below). From Eqs. (\ref{Eq9}) and (\ref{Eq11}) one finds that the
time-independent function $\beta(x_0,q)$ satisfies the compact
relation
\begin{equation}\label{Eq12}
\beta^{x_0+1}-\beta^{x_0}+q^{x_0}(1-q)=0\  .
\end{equation}
In particular, the asymptotic large-$t$ behavior of the survival
probability function $S(t;x_0,q)$ for the biased Sisyphus random
walkers is determined by the largest positive root of the polynomial
equation (\ref{Eq12}). This equation can be solved analytically in
the regime $x_0\gg1$ [$\beta(x_0,q)\simeq1$], in which case one
finds
\begin{equation}\label{Eq13}
\beta(x_0,q)=1-q^{x_0}(1-q)\cdot\{1+O[x_0q^{x_0}(1-q)]\}\  .
\end{equation}
The normalization factor $\alpha(x_0,q)$ in (\ref{Eq11}) can be
determined by the characteristic relation (\ref{Eq10}) of the
survival probability function. In particular, from Eqs.
(\ref{Eq10}), (\ref{Eq11}), and (\ref{Eq13}), one finds
\begin{equation}\label{Eq14}
\alpha(x_0,q)={{1-q^{x_0}}\over{[1-q^{x_0}(1-q)]^{x_0}}}\  .
\end{equation}

Interestingly, one finds from Eqs. (\ref{Eq11}) and (\ref{Eq13})
that the discrete function $S(t;x_0,q)$, which determines the
survival probabilities of the biased Sisyphus random walkers, is
characterized by the asymptotic large-$t$ behavior
\begin{equation}\label{Eq15}
{\cal R^{\text{asym}}}(x_0,q)\equiv
{{S(t+1;x_0,q)}\over{S(t;x_0,q)}}=1-q^{x_0}(1-q)\ \ \ \ \text{for}\
\ \ \ x_0,t\gg 1\  .
\end{equation}

\section{Numerical confirmation}

In the present section we shall use direct numerical computations in
order to confirm the validity of the analytically derived asymptotic
ratio (\ref{Eq15}) which characterizes the survival probability
function of the biased Sisyphus random walk model. In Table
\ref{Table1} we present the time-dependent ratio ${\cal
R}^{\text{exact}}(t)\equiv S(t+1;x_0=10,q=3/4)/S(t;x_0=10,q=3/4)$ as
computed directly from the recurrence relation (\ref{Eq9}) with the
boundary conditions (\ref{Eq3}) and (\ref{Eq10}). The data presented
in Table \ref{Table1} reveals the fact that the characteristic
dimensionless ratio ${\cal R}^{\text{exact}}(t)$ of the
time-dependent survival probabilities is described extremely well by
the analytically derived asymptotic ($x_0,t\gg 1$) ratio
(\ref{Eq15}) \cite{Notebtr}.

\begin{table}[htbp]
\centering
\begin{tabular}{|c|c|c|c|c|c|c|c|c|c|}
\hline \text{Time step $t$} & \ 12\ \ & \ 16\ \ & \
20\ \ & \ 24\ \ & \ 28\ \ & \ 32\ \ & \ 36\ \ & \ 40 \\
\hline \ ${\cal R}^{\text{exact}}(t)\equiv S(t+1)/S(t)$\ \ \ &\ \
0.9846\ \ \ &\ \ 0.9836\ \ \ &\ \ 0.9835\ \ \ &\ \ 0.9834\
\ \ &\ \ 0.9833\ \ \ &\ \ 0.9833\ \ \ &\ \ 0.9833\ \ \ &\ \ 0.9833\ \ \\
\hline
\end{tabular}
\caption{Survival probabilities in the biased Sisyphus random walk
model with an absorbing boundary. We present the time-dependent
dimensionless ratio ${\cal R}^{\text{exact}}(t)\equiv
S(t+1;x_0,q)/S(t;x_0,q)$ of the survival probability function as
computed directly from the recurrence relation (\ref{Eq9}). The data
presented is for the case $x_0=10$ with $q=3/4$. It is found that
the survival probability function, which characterizes the biased
Sisyphus random walk model with an absorbing trap, is described
extremely well by the analytically derived asymptotic relation
${\cal R}^{\text{asym}}=1-(1/4)\cdot(3/4)^{10}\simeq 0.9859$ [see
Eq. (\ref{Eq15})]. Note that the agreement between the analytical
and numerical results is even better if one uses the asymptotic
relation $S(t+1;x_0,q)/S(t;x_0,q)=\beta(x_0,q)$ [see Eq.
(\ref{Eq11})], where $\beta(x_0=10,q=3/4)=0.9833$ is determined
directly from the characteristic polynomial equation (\ref{Eq12}).}
\label{Table1}
\end{table}

\section{Asymptotic survival probabilities of the Sisyphus random walkers versus the standard random walkers}

In the present section we shall use the analytically derived results
of the previous sections in order to determine which type of a biased
random walker, the Sisyphus one [which is characterized by the
restart mechanism (\ref{Eq2})] or the standard one, has the larger
chance to survive for asymptotically long times.

Taking cognizance of Eqs. (\ref{Eq1}) and (\ref{Eq15}) one realizes
that, in the dimensionless regime $q>1/2$, the answer to the above
stated question is determined by the dimensionless ratio
\begin{equation}\label{Eq16}
{\cal F}(x_0,q)={{1-q^{x_0}(1-q)}\over{2\sqrt{q(1-q)}}}\ \ \ \ \text{for}\
\ \ \ x_0,t\gg 1\  .
\end{equation}
In particular, the biased Sisyphus random walkers are characterized
by a larger asymptotic survival probability (as compared to the
standard walkers) in the regime ${\cal F}(x_0,q)>1$, which
corresponds to the inequality
\begin{equation}\label{Eq17}
x_0>x^{\text{crit}}_0(q)\equiv
{{\ln\Big[{{1-2\sqrt{q(1-q)}}\over{1-q}}\Big]}\over{\ln(q)}}\ \
\Longleftrightarrow\ \
S^{\text{Sisyphus}}(t)>S^{\text{standard}}(t)\ \ \ \text{for}\ \ \
t\gg1\  .
\end{equation}
In Table \ref{Table2} we present explicit numerical values for the
$q$-dependent physical parameter $x^{\text{crit}}_0(q)$, which
determines the critical initial gap between the walkers and the trap
above which the late-time survival probability function of the
biased Sisyphus random walkers is {\it larger} than the
corresponding survival probability function of standard random
walkers with the same value of the biased jumping probability $q$.

\begin{table}[htbp]
\centering
\begin{tabular}{|c|c|c|c|c|c|c|c|c|c|c|}
\hline \text{\ Biased jumping probability $q$\ } & \ 0.501\ \ & \
0.51\ \ & \ 0.53\ \ & \ 0.55\ \ & \ 0.57\ \ & \ 0.6\ \ & \ 0.7\ \ & \ 0.8\ \ & \ 0.9\ \  \\
\hline \ \text{Critical initial gap}
$\left\lceil{x^{\text{crit}}_0(q)}\right\rceil$\ \ &\ \ 18 \ \ &\ \ 12\ \ \ &\ \ 9\ \ \ &\ \ 8\ \ \ &\ \ 7\ \ \ &\ \ 6\ \ \ &\ \ 4\ \ \ &\ \ 1\ \ \ &\ \ 1\ \  \\
\hline
\end{tabular}
\caption{The $q$-dependent critical initial gap
$x^{\text{crit}}_0(q)$ between the walkers and the trap. We display
the values $\left\lceil{x^{\text{crit}}_0(q)}\right\rceil$ of this
physical parameter for various values of the biased jumping
probability $q$ [here $\left\lceil{x}\right\rceil$ is the ceiling
function (the smallest integer which is larger than or equal to
$x$)]. The physical importance of this critical (marginal) physical
parameter stems from the fact that biased Sisyphus random walkers
with $x_0\geq x^{\text{crit}}_0(q)$ are characterized by late-time
survival probabilities which are {\it larger} than the corresponding
survival probabilities of standard biased random walkers with the
same value of the dimensionless physical parameter $q$.}
\label{Table2}
\end{table}

From the data presented in Table \ref{Table2} one learns that the
critical parameter $x^{\text{crit}}_0(q)$ is a monotonically
decreasing function of the biased jumping probability $q$. In
addition, it is interesting to note that the marginal gap
$x^{\text{crit}}_0(q)$ diverges in the $q\to 1/2$ limit. In
particular, from Eq. (\ref{Eq17}) one finds the asymptotic
functional behavior
\begin{equation}\label{Eq18}
x^{\text{crit}}_0(\epsilon)=-{{2\ln(\epsilon)}\over{\ln2}}-2+O[\epsilon\ln(\epsilon)]\
\ \ \text{for}\ \ \ \epsilon\equiv q-{1\over2}\ll1\
\end{equation}
for the critical (marginal) initial gap in the $q\to 1/2$ limit.

\section{Summary and Discussion}

In the present paper we have determined the time-dependent survival
probability function $S(t;x_0,q)$ of biased Sisyphus random walkers.
These walkers are characterized by the restart mechanism
(\ref{Eq2}), according to which each random walker has a finite
probability $1-q$ to return to her initial position $x_0$ at each
discrete time step (and a complementary probability $q>1/2$ to step
towards the absorbing trap at the origin).

It has been proved that the Sisyphus random walkers, like standard
biased random walkers, are characterized by exponentially decaying
late-time survival probabilities [see Eq. (\ref{Eq11})]. However, we
have explicitly shown that these two groups of biased random walkers
have different asymptotic decaying exponents [see Eqs. (\ref{Eq1})
and (\ref{Eq15})]. In particular, our analysis has revealed the
intriguing fact that, for a given value $q$ of the biased jumping
probability parameter, the survival probability function
$S(t;x_0,q)$ is characterized by a critical value
$x^{\text{crit}}_0(q)$ [see Eq. (\ref{Eq17})] of the initial gap
between the walkers and the trap, above which the asymptotic
survival probability of the biased Sisyphus random walkers is {\it
larger} than the corresponding survival probability function of
standard random walkers (who have no restart mechanism).

Finally, it is interesting to point out that, based on the fact that
the initial gap $x_0$ between the random walkers and the absorbing
trap at the origin is a positive integer with $\text{min}(x_0)=1$,
one finds from (\ref{Eq17}) that the biased Sisyphus random walkers
would necessarily (regardless of the value of $x_0$) have a larger
survival probability (as compared to the standard random walkers) in
the dimensionless regime
\begin{equation}\label{Eq19}
q>{1\over2}\Big(1+\sqrt{8\sqrt{2}-11}\Big)\simeq 0.78\
\end{equation}
of the biased jumping probability.

\bigskip
\noindent
{\bf ACKNOWLEDGMENTS}
\bigskip

This research is supported by the Carmel Science Foundation. I would
like to thank Yael Oren, Arbel M. Ongo, Ayelet B. Lata, and Alona B.
Tea for helpful discussions.

%\newpage


\begin{thebibliography}{99}

\bibitem{RW1} M. N. Barber and B. W. Ninham, {\it Random and Restricted
Walks} (Gordon and Breach, New York, 1970).

\bibitem{RW2} N. G. van Kampen, {\it Stochastic Processes in Physics and
Chemistry} (North-Holland, Amsterdam, 1992).

\bibitem{RW3} R. Fernandez, J. Frohlich, and A. D. Sokal, {\it Random Walks,
Critical Phenomena, and Triviality in Quantum Field Theory}
(Springer Verlag, Berlin, 1992).

\bibitem{RW4} G. H. Weiss, {\it Aspects and Applications of the Random Walk} (North
Holland, Amsterdam, 1994).

\bibitem{RW5} D. ben-Avraham and S. Havlin, {\it Diffusion and Reactions in Fractals
and Disordered Systems} (Cambridge University Press, Cambridge,
2000).

\bibitem{RW6} R. Dickman and D. ben-Avraham, Phys. Rev. E. {\bf 64}, 020102(R) (2001).

\bibitem{DV1} J. Marro and R. Dickman, {\it Nonequilibrium Phase Transitions
in Lattice Models}, (Cambridge University Press, Cambridge, 1999);
H. Hinrichsen, Adv. Phys. {\bf 49}, 815 (2000).

\bibitem{DV2} S. Hod, Phys. Rev. Lett. {\bf 90}, 128701 (2003)
[arXiv:cond-mat/0212055]; S. Hod, Phys. Rev. Lett. {\bf 105}, 208701
(2010) [arXiv:1009.0941].

\bibitem{DV3} Z. Farkasa and T. Fulop, J. Phys. A: Math. Gen. {\bf 34}, 3191 (2001).

\bibitem{DV4} K. De'Bell and T. Lookman, Rev. Mod. Phys. {\bf 65}, 87 (1993).

\bibitem{DV5} P. Grassberger, H. Chate, and G. Rousseau, Phys. Rev. E {\bf55}, 2488 (1997).

\bibitem{Notesur} The survival function $S(t)$ is defined as the time-dependent
probability that a random walker has not been absorbed by the trap
after taking $t$ steps.

\bibitem{Sisy} M. Montero and J. Villarroel, Phys. Rev. E {\bf 94}, 032132 (2016).

\bibitem{Hodanp} S. Hod, Annals of Phys. {\bf 406}, 200 (2019).

\bibitem{RS1} M. R. Evans and S. N. Majumdar, Phys. Rev. Lett. {\bf 106}, 160601 (2011).

\bibitem{RS2} D. Boyer and C. Solis-Salas, Phys. Rev. Lett. {\bf 112}, 240601 (2014).

\bibitem{RS3} L. Kusmierz, S. N. Majumdar, S. Sabhapandit, and G.
Schehr, Phys. Rev. Lett. {\bf 113}, 220602 (2014).

\bibitem{RS4} X. Durang, M. Henkel, and H. Park, J. Phys. A {\bf 47}, 045002 (2014).

\bibitem{RS5} D. Boyer and I. Pineda, Phys. Rev. E {\bf 93}, 022103 (2016).

\bibitem{Notextr} We assume, without loss of generality, that the
absorbing trap is located at the origin $x_{\text{trap}}=0$.

\bibitem{Notefct1} This relation follows from
the jumping rule (\ref{Eq2}), according to which a biased random
walker who is located at $x_0$ has to go $x_0$ steps in a row
towards the trap (with a characteristic probability $q$ for each
step to the left) in order to be absorbed at the origin.

\bibitem{Notefct2} Here we have used the jumping rule (\ref{Eq2}), according to which a
biased Sisyphus random walker has a probability of $1-q$ to jump
from her current position $x(t)$ back to her initial position $x_0$.

\bibitem{Notebtr} It is worth noting that the exact ratio ${\cal R}^{\text{exact}}(t)$ agrees even
better with the asymptotic relation
$S(t+1;x_0,q)/S(t;x_0,q)=\beta(x_0,q)$ [see Eq. (\ref{Eq11})], where
the function $\beta(x_0,q)$ is determined numerically from the
polynomial equation (\ref{Eq12}).

\end{thebibliography}
\end{document}